\documentstyle [twocolumn,aps]{revtex}%

\begin{document}
\draft

\title{On the difference between proton and neutron spin-orbit splittings
in nuclei}

\author{V.I.\thinspace Isakov$^1$, K.I.\thinspace Erokhina$^2$,
H.\thinspace Mach$^3$, M.\thinspace Sanchez-Vega$^3$,
B.\thinspace Fogelberg$^3$}

\address{$^1$ Petersburg Nuclear Physics Institute,
Russian Academy of Sciences, Gatchina 188300, Russia}

\address{$^2$ Physicotechnical Institute,
Russian Academy of Sciences, St.Petersburg 194021, Russia}

\address{$^3$ Department of Radiation Sciences, Uppsala University,
 Nyk\"{o}ping S-61162, Sweden}

\date{\today}%
\maketitle

\begin{abstract}%
The latest experimental data on nuclei at $^{132}$Sn permit us for
the first time to determine the spin-orbit splittings of neutrons and
protons in identical orbits in this neutron-rich doubly-magic
region and compare the case to that of $^{208}$Pb.
Using the new results, which are now consistent for the two neutron-rich
doubly magic regions, a theoretical analysis defines
the isotopic dependence of the mean field spin-orbit potential and leads
to a simple explicit expression for the difference between the
spin-orbit splittings of neutrons and protons. The isotopic dependence is
explained in the framework of different theoretical approaches.

\end{abstract}%
\narrowtext

\section{Introduction}

Spin-orbit splitting of the mean field orbitals is one of the main
factors, which determine nuclear structure in nuclei both near and far
from the closed shells. While the global characteristics of spin-orbit
splitting are well known, one cannot say the same about the isotopical
dependence of splitting. However,
the new experimental results obtained recently \cite{1,2,3} on nuclei
close to $^{132}$Sn allowed to define a nearly complete set of neutron
and proton single-particle orbitals and some important statical and
dynamical properties of the mentioned nuclides.
In particular, from the measurements on $^{133}$Sb new information was
obtained \cite{2} on the energies and the decay properties of proton
single particle states above the $Z=50, N=82$ shells. One
purpose of our study is to evaluate the new results on the single
particle levels at $^{132}$Sn and intercompare them to the known data on
such states in the other doubly closed shell (DCS) regions. However, our
main aim is to examine the magnitude of the spin-orbit splittings of
neutrons and protons in identical orbits, and to determine their isospin
dependence. A preliminary account of this work was given in
\cite{isa2000}.

The determination of the isospin dependence has a broader significance,
since the magnitude of the spin-orbit splittings could be one of
the factors contributing
to significant structural changes in nuclides having an extreme neutron
excess. Consequently, the trends extracted from the empirical data
presently available are crucial guidelines for the theoretical analysis,
and it is important to show that model calculations indeed do
reproduce such trends.

The following presentation begins with an analysis
of the existing experimental data in Section II, followed by
theoretical considerations and evaluations in Section III. A
summary and conclusions are included in the last section.

\section{Experimental Values}

We now examine the experimental data on the spin-orbit splitting
for neutrons and protons for two groups of the doubly closed shell
nuclei: the $N=Z$ and the neutron-rich regions. In order to
facilitate the comparison, the available systematics of single particle
energies at the $N=Z$ nuclei of $^{16}$O, $^{40}$Ca, $^{100}$Sn, and the
neutron-rich $^{132}$Sn and $^{208}$Pb are presented in Tables 1 to 5,
respectively. Here, the energies of the particle and hole states closest
to the Fermi level were determined from the differences of binding
energies of the core and the corresponding adjacent odd nuclei:
$\varepsilon$(particle)=B(core)--B(core+nucleon) and
$\varepsilon$(hole)=B(core--nucleon)--B(core), using the
experimental binding energies from \cite{20}. The energies of orbitals
more distant from the Fermi-level were subsequently defined by the
addition (subtraction) of the experimentally determined excitation
energies \cite{1,2,15,16,17,18,21,22} of the corresponding orbitals
in the adjacent odd nuclei. Important for this process is that we have
accounted for the fragmentation of states in the cases when the
pertinent data were available. The cases where the effect is
essential are indicated in Tables 1$-$5
by an asterisk (*) next to the experimental value.

\subsection{N=Z doubly-closed shell regions}

The spin-orbit splittings in the $N=Z$ nuclei can be determined using the
data in Tables 1--3. Here the spin-orbit splittings of the
$1p_{1/2}$$-$$1p_{3/2}$ and $1d_{3/2}$$-$$1d_{5/2}$ orbits in $^{16}$O
\cite{28,29} are practically equal for protons and neutrons, with the
values of 6.32 and 6.17 MeV (the difference being $-$2.4$\%$) and 5.00 and
5.08 MeV (the difference being +1.6$\%$). Similarly, for $^{40}$Ca,
Ref.~\cite{30}, we have
the values for protons and neutrons equal to 2.01 and 2.00 MeV for the
$2p_{1/2}$$-$$2p_{3/2}$ orbit, 6.00 and 6.00 MeV for the
$1d_{3/2}$$-$$1d_{5/2}$ orbit, and 4.95 and 4.88 MeV for the
$1f_{5/2}$$-$$1f_{7/2}$ case.
In the absence of experimental data on the single particle states at
$^{100}$Sn, we adopt here the extrapolated single particle energies from
Grawe {\it et al.} \cite{31}, as is shown in Table 3. Based on these data,
one may conclude that, within the errors, the spin-orbit splitting of
the $1g_{7/2}$$-$$1g_{9/2}$ orbit is also equal for protons and neutrons,
namely, at 6.82(28) and 7.00(28) MeV. These six cases show that the
splitting for the $N=Z$ DCS regions is practically equal, with small
oscillations either way, but below 2.6$\%$. This equality simply reflects
the concept of isobaric invariance in nuclei.

\subsection{Neutron-rich doubly-closed shell regions}

For the neutron-rich nuclei we find the situation considerably
different, see Tables 4--5 for the data, and Table 6 regarding
the splittings.
In the $^{132}$Sn region, the energy of the $3/2^+$ proton state in
$^{133}$Sb was recently determined \cite{2} at 2.44 MeV. Using this
value and the previously determined single particle excitations in
nuclei close to $^{132}$Sn (see \cite{1,15,16,17,18}) the spin-orbit
splittings of the $2d$ levels both in proton and neutron systems at
$^{132}$Sn can be now defined. The $2d_{3/2}$$-$$2d_{5/2}$ splitting was
found to be 1.48 MeV for protons and 1.65 MeV for neutrons. This means
that the neutron spin-orbit splitting is somewhat larger (by more than
11$\%$) than for protons.

In the case of $^{208}$Pb, it was noted \cite{2} that the situation
seemed to be quite opposite. Namely, a simple analysis of the single
particle levels in $^{209}$Bi and $^{207}$Pb suggested \cite{2}
that the spin-orbit splitting of the $2f_{5/2}$$-$$2f_{7/2}$
orbit is equal to 1.93 MeV for protons and 1.77 MeV for neutrons.
However, a significant correction is needed. It follows from the
experimental evidence that the neutron $2f_{7/2}$ state in $^{207}$Pb is
strongly fragmented, while the conclusions in Ref.~\cite{2} were derived
by considering only the lowest, albeit the strongest component of this
state. In order to identify, in the spirit of Refs.~\cite{Cohen63,Bar70},
the true single particle energy of the neutron $2f_{7/2}$ state,
we use the weighted average of the fragmented $7/2^-$ energy levels, with
the weight provided by the spectroscopic factors determined in the (d,t)
reaction on $^{208}$Pb \cite{19}. In this way we obtain a more accurate
unfragmented excitation energy of this state equal to 2.70 MeV (instead
of 2.34 MeV).  Using this excitation energy, included in Table 5, we
find the neutron spin-orbit splitting of the $2f$ orbit as 2.13 MeV,
which, similarly to the case of the $2d$ orbit in $^{132}$Sn, is larger
by about 10$\%$ than the splitting of 1.93 MeV for protons.

An additional piece of evidence along the same line is given by the
analysis of the $3p_{1/2}$$-$$3p_{3/2}$ spin-orbit splitting at $^{208}$Pb.
One obtains 0.85 MeV for protons (after
correcting for the fragmentation of
the proton $3p_{1/2}$ level) and 0.90 MeV for neutrons. Thus again the
value for neutrons is larger by about 6$\%$ than for protons.
Consequently, based on the three cases described above, it is evident
that the neutron spin-orbit splitting in the neutron-rich DCS nuclei of
$^{208}$Pb and $^{132}$Sn, is {\it systematically larger by
$\sim$10$\%$} than the corresponding proton splitting.

\subsection{Fragmentation of strength at $^{132}$Sn}

The fragmentation of single particle states at $^{208}$Pb is
mainly influenced by the presence of a very low-lying and highly-collective
$3^{-}_{1}$ phonon state at 2.62 MeV. The effects caused by the $2^{+}_{1}$
state are less as the collectivization of quadrupole phonon, and the
corresponding nucleon-phonon vertexes are small in heavy nuclei near doubly
closed shells. For example, in the neutron "hole"
$^{207}$Pb nuclei, the main part of energy shift of the $7/2^{-}$ level
is caused by mixing with the higher lying
$(3^{-}_{1} \otimes {\nu1i_{13/2}}^{-1})_{7/2^{-}}$ state.
Numerical evaluation performed by using the quasiparticle--phonon model
with the coupling constant extracted from the $B(E3; 3^-_1 \to$ ground
state$)$ value shows that the  $7/2^{-}$ level corresponding to
the "pure"  ${\nu2f_{7/2}}^{-1}$ state moves down by the amount of $\sim$
0.4 MeV, thus approaching the experimental value of 2.34 MeV.
This large shift is due to rather strong coupling constant and non
spin-flip nature of the matrix element. The magnitude of the predicted
shift is very close to the experimental value of $2.70 - 2.34 = 0.36$ MeV
mentioned above. At the same time,
the $(3^{-}_{1} \otimes {\nu1i_{13/2}}^{-1})$ configuration has no
$5/2^{-}$ component and thus one does not observe experimentally the
fragmentation of the lower lying $5/2^{-}$ level at 0.57 MeV. We note
here that the $7/2^{-}$ and the $5/2^{-}$ states in the proton
"particle" $^{209}$Bi nucleus have the opposite ordering,
$7/2^{-}$ being the lower one. Due to mixing with the
$(3^{-}_{1} \otimes \pi1i_{13/2})$ configuration the "pure" $\pi2f_{7/2}$
level is also pushed down, but only by about 0.2 MeV due to larger
energy difference. Thus, after taking account of configuration mixing,
not only the neutron $\Delta^{(n)}_{\ell s}(2f)$ splitting between
the pure states increased as compared to 1.77 MeV, but also the  proton
$\Delta^{(p)}_{\ell s}(2f)$ splitting decreased to a smaller value.

Turning to the region of $^{132}$Sn, we note that the corresponding
experimental data on fragmentation of single particle states are not known
at present. However, as was pointed out by Blomqvist \cite{Blom81}, the
$^{132}$Sn and
$^{208}$Pb nuclei are in some respect twins, having similar shell
structures with the correspondence of $ l \to l + 1,\, j \to j + 1 $
for most of the orbitals in these regions. Therefore, all the arguments
presented above for splitting of the $2f$ levels at $^{208}$Pb are
completely valid also for the $2d$ states at $^{132}$Sn,
with replacement of $1i_{13/2}$ by $1h_{11/2}$. So far there is no
direct experimental data on the $B(E3; 3^-_1 \to$ ground state$)$ value
in $^{132}$Sn. However, the core has much higher rigidity here in
comparison with $^{208}$Pb and the energy of the $3^{-}_{1}$ state is
substantially higher at 4.35 MeV. Thus from accounting for
configuration mixing one expects some further increase of the
$\Delta ^{(n)}_{\ell s}(2d)$ splitting and a decrease of
$\Delta^{(p)}_{\ell s}(2d)$, but these changes should be smaller than
for the $2f$ levels at $^{208}$Pb.
Estimates based on an indirect evaluation of the $B(E3)$ value from
the magnitude of the octupole effective charge in $^{134}$Te \cite{Omtv95}
confirm the pattern of changes of the $\Delta^{(p,n)}_{\ell s}(2d)$
values presented above. However, in the absence of
experimental data on direct reactions we present in Tables 4 and 6
the values of energies at $^{132}$Sn that do not include
averaging over spectroscopic factors.

\section{Theoretical Approach}

\subsection{General considerations}

Turning to the theoretical interpretation \cite{isa2000} of the
experimental values of the spin-orbit splitting discussed above, we
shall first recall that from the point of view of many-body theory the
average spin-orbit potential has its origin in the
pair spin-orbit interaction between nucleons (with
tensor forces providing a minor contribution as well).
On the level of qualitative
arguments it was noted by Bohr and Mottelson \cite{32} that due
to the symmetry properties one should expect the neutron
spin-orbit splitting to be somewhat larger than that for protons
in heavier nuclei, simply due to a higher number of like
particles in the neutron case.
However, at that time the absence of experimental data did
not permit a meaningful comparison with
measurements. With the presently available
data we can fill this gap, providing also some quantitative
considerations.

The two-body spin-orbit interaction differs from zero only in the states
with a total spin $S=1$. The neutron-neutron and proton-proton
systems have the total isospin $T=1$ and thus due to the Pauli
principle have odd values of the relative orbital momentum $L$
(in fact, $L=1$). At the same time, the neutron-proton system is
composed from the $T=0$ and $T=1$ states with equal weights,
having $L=0$ and $L=1$, correspondingly. Due to the absence of
spin-orbit interaction in states with $L=0$, the pair spin-orbit $np$
interaction is half as strong as that in $pp$ or $nn$-systems.

If $U_{\ell s}(n)$ and $U_{\ell s}(p)$  represent
the magnitudes of the mean spin-orbit field for neutrons and
protons and $\vartheta(T=1, S=1, L=1)$
is a quantity representing the parameter of the
pair spin-orbit interaction in a state with $T=1, S=1, L=1$ then
the above discourse leads to $$U_{\ell s}(n)\sim \vartheta(1, 1, 1)
\cdot\left(N+\frac 12 Z\right) \equiv \vartheta \cdot\left(A-\frac
Z2\right) \, {\rm and}$$ $$ U_{\ell s}(p)\sim \vartheta(1, 1, 1)\cdot
\left(\frac N2+Z\right) \equiv \vartheta \cdot\left(A-\frac N2\right)
.\eqno{(1)}$$

As the spin-orbit splitting $\Delta_{\ell s}^{(n,p)}\sim
U_{\ell s}(n,p)$, the relative {\it difference} "$\varepsilon$" of the
neutron and proton spin-orbit splittings is given by the expression:

$$\varepsilon=\frac{\Delta_{\ell s}^{(n)}-\Delta_{\ell s}^{(p)}}
{(\Delta_{\ell s}^{(n)}+\Delta_{\ell s}^{(p)})/2}
=\frac 23\,
\frac{N-Z}{A}\,.\eqno{(2)}$$

On the other hand, we can express the strength of the
spin-orbit mean field in the form:

$$U_{\ell s}(\tau_{3})=V_{\ell s}\left(1+\frac{1}{2}\,
\beta_{\ell s}\frac{N-Z}{A}\cdot
\tau_{3}\right)\,.\eqno{(3)}$$

\noindent
Here $\tau_{3}=-1$ for neutrons, $\tau_{3}=+1$ for protons
and $\beta_{\ell s}$ is the parameter  that defines the isospin
dependence of the mean spin-orbit field. Then we easily obtain, this
time in terms of eq.~(3), an expression for the relative
difference between the spin-orbit splittings of neutrons and protons
in identical orbits, $\varepsilon$:

$$\varepsilon=-\beta_{\ell s}\frac{N-Z}{A}\,.\eqno{(4)}$$

\noindent
It follows from a comparison of eqs.~(2) and (4) that
$\beta_{\ell s}=-2/3$.

Strictly speaking, this derivation was performed for the two-body
spin-orbit interaction. However, as mentioned above, tensor forces provide
also some contribution to the spin-orbit splitting. This non-central
interaction is proportional to $S_{12}$ with

$$S_{12}=3(\bf{\sigma}_{1} \bf{n})(\bf{\sigma}_{2} \bf{n}) -
\bf{\sigma}_{1} \bf{\sigma}_{2} = $$
$$ =\sqrt{24 \pi} \cdot [[\sigma _{1} \otimes
\sigma_{2}]^{2} \otimes Y_{2}]^{0}_{0}. \eqno{(5)}$$

One can easily see from (5) that the diagonal matrix elements of this
interaction are different from zero only for states with $S = 1$
and $L \geq 1$, of which the $S = T = L = 1$ one is of the main
importance. It is just the state which was already considered
in this subsection in the case of spin-orbit interaction.
Consequently, the diagonal part of tensor forces also provides
contribution of the type given by eq.~(3) with $\beta_{\ell s}
= - 2/3$, and thus it leads only to a renormalization of the $V_{\ell s}$
value. However, as the spatial part of tensor operator is proportional
to $Y_{2}(\bf n)$ and due to the spin structure of $S_{12}$, this
renormalization equals zero in cases of spin saturated
spherical nuclei. Thus in  $^{16}$O and $^{40}$Ca tensor forces give
a contribution to the isoscalar part of the spin-orbit splitting, that is
mediated by their non-diagonal part and caused by admixtures, that are out
of the Hartree--Fock type ground state. As was shown in Ref.~\cite{Pieper93},
tensor forces may really lead to a substantial contribution
to the isoscalar part of spin-orbit splitting. At the same time,
in nuclei that are not spin saturated, such as $^{48}$Ca, tensor forces
can contribute to the spin-orbit splitting even in the "diagonal" scheme
(i.e.: a scheme without admixtures), if the antisymmetrization is
properly included. Our numerical calculations for seniority one states
of $^{47}$Ca and $^{47}$K both having one neutron or proton hole and
performed in the framework of the multiparticle shell model with tensor
forces taken from our previous works \cite{12}--\cite{11}, have
demonstrated that the inclusion of a tensor component of the interaction
leads to energy shifts that correspond to some variation of the spin-orbit
splittings $\Delta_{\ell s}$, such that in $^{48}$Ca
$\Delta^{(n)}_{\ell s}(1d) - \Delta^{(p)}_{\ell s}(1d) = 0.34$ MeV and
$\Delta^{(n)}_{\ell s}(1p) - \Delta^{(p)}_{\ell s}(1p) = 0.24$ MeV.
These shifts arise from neutrons filling the $\nu1f_{7/2}$ subshell and
are mainly due to charge exchange two-body matrix elements of the
$np$-interaction mediated by the isovector part of the tensor force
$(\sim \tau _{1} \tau_{2})$. Thus the inclusion of tensor forces does not
change the pattern of spin-orbit splitting, which also leads to negative
values of $\beta_{\ell s}$ ranging from about $-$0.4 to $-$0.7. These
results qualitatively agree with those presented in Ref.~\cite{Davies71},
where in the framework of the Brueckner--Hartree--Fock method with
Reid potential (containing both the spin-orbit and tensor components),
a substantially larger neutron than proton splitting was obtained for
the $1p$ and $1d$ orbitals in $^{48}$Ca with $\beta_{\ell s}$ in the
range from about $-$0.5 to  $-$1.8. We note that the data on spin-orbit
splittings of the $2d$ states in $^{132}$Sn as well as on the splittings
of the $2f$ and $3p$ levels in $^{208}$Pb lead to effective values of
$\beta_{\ell s}$ equal to $-0.55$, $-0.60$ and $-0.27$, respectively,
which are numbers in very satisfactory general agreement with the
prediction of eq.~(4).

It is thus of substantial interest to evaluate to what
extent the isotopic dependence of the spin-orbit splittings are
reproduced by standard model calculations. Three different
approaches were made as described below.

\subsection{Evaluation I: Walecka model}

The first evaluation is made in the Hartree approximation starting from
the Dirac phenomenology with meson-nucleon interactions according
to the Walecka model \cite{Wal74}.
One obtains (see for example \cite{Brock78}--\cite{Yosh99} and
references therein) a Skyrme-type single particle equation for
a nucleon having the effective mass $m_{N}^{*}$. This approach well
explains the magnitude of spin-orbit splitting in nuclei.
Here, and mainly for heavier nuclei, we concentrate only on the
{\it difference} between the proton and neutron splittings of spin-orbit
partners in the same nuclei and
resulting from a spin-orbit potential having the form
(see for example \cite{Rein86}--\cite{Bir98}):

$$\hat{U}_{\ell s}= \frac{\lambda_{N}^{2}}{2}\, \frac{1}{r}\,
\{(\frac{m_{N}}{m_{N}^{*}})^2 \frac{d}{dr}\lbrack ( V_{\omega}^{0} -
S_{\sigma,{\sigma}_0}^{0})-\,$$
$$- (V_{\rho}^{1}-S_{\delta,\sigma,{\sigma}_0}^{1})\cdot
\tau_{3} \rbrack -2k(\frac{m_{N}}{m_{N}^{*}})
\frac{d}{dr} V_{\rho}^{1} \cdot \tau_{3} \}\,
\hat{\bf \ell} \cdot \hat{\bf s}, \, \eqno{(6)}$$

\noindent
where $V = V^{0}-\tau_{3}\cdot V^{1}$ \,and \,
$S = S^{0} -\tau_{3}\cdot S^{1}$ are the vector and
scalar fields related to corresponding mesons,
$m_{N}^{*} = m_{N} + \frac{1}{2}(S-V)$, while $k$ is the
ratio of tensor to vector coupling constants of $\rho$-meson.
Various approaches have been used to determine the coupling
constants.
In \cite{Bir98} the meson-nucleon coupling constants, defining
the $V$ and $S$ fields, were taken from the Bonn
$NN$ boson exchange potential \cite{Mac87}, where $\sigma$ and
${\sigma}_0$ are scalar mesons imitating the 2$\pi$ exchange
in the $NN$- systems with $T$=1 and $T$=0, correspondingly.
In other works  (see for example
\cite{Rein86}--\cite{Koepf91}) the constants were
defined from the description of global nuclear properties,
with inclusion of the $\sigma^{3}$ and $\sigma^{4}$ terms in
the Lagrangian density (one $\sigma$-meson with the same
characteristics for $T$=1 and $T$=0 channels was used, which
leads to zero contribution of this meson to  $S^1$
in formula (6); note also that the tensor term was not included in the
$\rho$-meson vertex in Refs. \cite{Rein86}--\cite{Koepf91}).

Taking into account
that the radial dependence of the $(m_{N}/m_{N}^{*})$ is much
weaker than that of $V$ and $S$, which are considered to be
proportional to the density in the form of Fermi function,
one can approximately present formula (6) as follows:

$$\frac{1}{r}\, \frac{df}{dr} \cdot V_{\ell s} \left( 1+
\frac{1}{2}\,\beta_{\ell s} \frac{N-Z}{A} \cdot\tau_{3} \right)
\hat{\bf \ell} \cdot \hat{\bf s} \,;$$
$$f =\lbrack 1+exp\,(\frac{r-R}{a}) \rbrack^{-1} \,. \eqno{(7)}$$

Calculating the $V$ and $S$ magnitudes in the
center of nuclei at the values of vector and scalar densities
${\rho}_{v}$ = 0.17, ${\rho}_{s}$ = 0.16,\, ${\rho}_{v}^{-}$ =
0.17\,$(N-Z)/A$, ${\rho}_{s}^{-}$ = 0.16\,
$(N-Z)/A$ (all in {fm}$^{-3}$) , using the coupling parameters
from \cite{Bir98}, \cite{Mac87} and taking into
account the isotopic dependence of $m_{N}/m_{N}^{*}$, we obtain
$V_{\ell s}$ $\approx$ 34 MeV$\cdot$ {fm}$^2$ and \, $\beta_{\ell s}$
$\approx$ -- 0.40.
If we use the NL2 set of parameters from
\cite{Lee86,Koepf91} then we have $V_{\ell s}$ $\approx$ 31 MeV $\cdot$
{fm}$^2$, $\beta_{\ell s}$ $\approx$ -- 0.43. At the same time the set NL1
from \cite{Rein86,Koepf91}, giving small values of effective masses,
leads to $V_{\ell s}$ $\sim$ 50 MeV $\cdot$ {fm}$^2$ and
$\beta_{\ell s}$ $\sim$ -- 1.3. As the $V^{1}, S^{1}$
magnitudes are proportional to ${\rho}_{v}^{-}$ and ${\rho}_{s}^{-}$,
both the formulae (6) and (7) give equal spin-orbit splitting
for protons and neutrons in the $N=Z$ nuclei. It should be noted,
that the value of
$\beta_{\ell s}$ is always negative and is determined mainly, or
entirely, by the $\rho$-meson conribution.

The magnitudes of the empirical effective values of $\beta_{\ell s}$
at $^{132}$Sn and $^{208}$Pb, listed in subsection IIIA, are quite well
reproduced by the model calculations in this subsection, in particular
by those using the first two sets of parameters.

It is worth mentioning that a study of the neutron spin-orbit splitting
in light nuclei as a function of $A$ at given $Z$ was recently performed in
the framework of the Walecka model by Lalazissis {\it et al.} \cite{Lala98}.
However, the intercomparison between the
splittings of both proton and neutron "similar" spin-orbit doublets in the
same nuclei was not performed there.

\subsection{Evaluation II: Woods-Saxon model}

In the second approach, using a Woods-Saxon (W-S) model, we
let the single particle levels be generated by the potential
$$\hat{U}(r,\hat \sigma,\tau_{3})=U_{0}(\tau_{3})f(r)\,+$$
$$+\, \frac{U_{\ell s}(\tau_{3})}{r}\,\frac{df}{dr}\,
\hat{\bf \ell} \cdot \hat{\bf s} +
\frac{(1+\tau_{3})}{2}\,U_{Coul}\,, \eqno{(8)}$$

\noindent
where $U_{0}(\tau_{3})=
V_{0}(1+\frac{1}{2}\beta\frac{N-Z}{A}\cdot \tau_{3})$;
$U_{\ell s}$ and $f(r,a,R)$ are defined by eqs.~(3) and (7),
$R=r_{0}A^{1/3}$, while $U_{Coul}(r,R_{c},Z)$  represents  the
potential of a uniformly charged sphere with the charge $Z$
and radius $R_{c}=r_{c}A^{1/3}$.

In previous works \cite{12} --\cite{11}, calculations were made
using the $V_{0}=-51.5$ MeV, $r_{0}$ =
1.27 fm, $V_{\ell s}$ =33.2 MeV $\cdot$ {fm}$^2$, $a(p)$= 0.67 fm, $a(n)$
= 0.55 fm and $\beta_{\ell s}$ =$\beta$ =1.39, which on the average described
the spectra of single particle states in nuclei from $^{16}$O
to $^{208}$Pb. This set of parameters is denoted here as the "Standard"
one. With the appearance of new experimental data on the
single-particle levels, we performed a new determination of parameter
values through the Nelder--Mead
method \cite{Nel67} by minimizing the root-mean square
deviation
$$\delta=\sqrt{\frac 1n\sum\limits_{k=1}\limits^{n}
(\varepsilon_k^{\rm theor}-\varepsilon_k^{\rm exp})^2}\,.
\eqno{(9)}$$

The computation demonstrated a very small sensitivity of results
to the value of $r_c$, which was adopted to be the same as before:
$r_c=1.25$ fm. The minimization of $\delta$ performed for
all nuclei presented in Tables 1--5 with $r_c=1.25$ fm and
different values of $r_0$, showed that the minimum in all cases
corresponds to $r_0\approx 1.27$ fm, that also coincides with
the value adopted by us before. The values $r_c=1.25$ fm and
$r_0=1.27$ fm were thus fixed in further calculations.

As was noted above, the optimal relation of proton to neutron
spin-orbit splitting corresponds to $\beta_{\ell s}\sim-0.6$.
The fourth column, "Set 1", of Tables 4 and 5
presents the values of theoretical energy levels obtained
in the optimization with fixed values of $\beta_{\ell s}=-0.6$,
$a_p=0.67$ fm and $a_n=0.55$ fm.

The fifth column, "Set 2", of Tables 4 and 5 presents the
results of optimization with only two fixed parameters: $a_p=0.67$ fm
and $a_n=0.55$ fm.

The values of "Set 3" corresponds to an optimization at fixed $\beta_{\ell
s}=-0.6$, while "Set 4" are the results with no parameters
fixed.

We see that the optimized values of $V_0$, $V_{\ell s}$ and
$\beta$ (see formula (8)) are very close to the "Standard"
ones, with small variations from nucleus to nucleus. The
magnitudes of the diffusinesses "$a$" vary more strongly,
differing by about 10$\%$ to 15$\%$ from their "standard" values.
A comparison of the "Stnd" with "Set 1"
and of "Set 3" with "Set 4" results shows that the contribution of
$\beta_{\ell s}$ to the root-mean square deviation $\delta$ is small.
It is thus more
reasonable to define $\beta_{\ell s}$ not from a minimization of
$\delta$, but rather by using the experimental and theoretical
arguments mentioned above.
This conclusion is confirmed by the results of
Koura and Yamada\cite{Koura20}, who made a number of different fits
of W-S parameters to the same set of experimental data, obtaining
diverse (in magnitude and sign) values of  the  parameter that
defines the  contribution to the spin-orbit term, which is linear
in $(N-Z)/A$. A global adjustment of W-S parameters simply appears
to be only weakly sensitive to details of the spin-orbit splitting.

As mentioned previously,
the energies of levels in nuclei with $N=Z$ (see Tables 1--3)
are independent of $\beta$ and $\beta_{\ell s}$. Here the optimization
was performed twice, first with fixed values of $a_n=0.55$ fm
and $a_p=0.67$ fm with a subsequent definition of $V$ and $V_{\ell s}$
("Set 1") and secondly without fixing any parameters ("Set 3").

The results of the calculations presented in Tables 1 to 5 include
some levels having positive energies, i.e. unbound but
sub-barrier states. In such cases we present here the real
part of the single particle energies only for those states having very
small decay widths.

To summarize Evaluation II, we have determined the parameters of the
W-S potential using a global mean square-root optimization, except for
the isospin dependent spin-orbit term, where the parameter value was
found to be insensitive to the adjustment. Hence the value of
$\beta_{\ell s}$$\sim$$-$0.6 was deduced from physical considerations
based on experimental spin-orbit splittings.

\subsection{Evaluation III: Hartree-Fock with a Skyrme interaction}

For the third model approach, which complements the first two
evaluations using the empirically adjusted W-S potential (8) and the
microscopical procedure, we have selected the Hartree-Fock
calculations with the SIII interaction. The results of these
self-consistent calculations, listed in the last two columns
of Tables 1 to 5, were obtained by considering
the contribution of a single-particle part of the center-of-mass
energy and taking into account the Coulomb exchange term in the
Slater approximation. The SIII-1 results correspond to calculations
which take into account all terms of the energy functional
contributing to spin-orbit splitting, while the SIII-2 results have
been obtained by omitting the spin density terms in the spin-orbit
potential. In the last case our results are close to those from the
study by Leander {\it et al.} \cite{37}
performed for $^{208}$Pb, $^{132}$Sn
and $^{100}$Sn nuclei. We see that the results obtained in the
framework of the Hartree-Fock method also demonstrate that the
calculated neutron spin-orbit splittings of the $2d$ orbit in
$^{132}$Sn as well as of the $2f$ and $3p$ orbits in $^{208}$Pb are
larger than for protons and they correspond to effective $\beta_{\ell
s}$ in the interval of $-$0.9 to $-$0.6.  We note that the difference
between the neutron and proton spin-orbit splittings is reproduced here
by using a simple parameterization of Skyrme forces. Our calculated
results differ from those of Noble \cite{Nob79} who proposed that the
isotopic dependence of the spin-orbit potential in the Hartree scheme
is cancelled through the contribution of exchange terms, but agree with
that of \cite{Davies71}. We mention here that the SIII parameterization
contains density-dependent terms that imitate in some sense  the
three-body interaction.

Tables 1 and 2 give results for $^{16}$O and $^{40}$Ca, which
are spin-saturated nuclei.
In these cases the spin density terms, included in
SIII-1 but not in SIII-2, do not contribute significantly to the
spin-orbit splitting (the contributions in these cases are only due
to small differences in the radial wave functions of spin-orbit partners).
Consequently, as can be expected, the SIII-1 and SIII-2 calculations
give very similar results in both cases.

\section{Discussion and conclusions}

Using theoretical analysis and systematics of available
experimental data we have derived formula (3) that describes
the {\it difference} between the neutron and proton
spin-orbit splittings, i.e. the isotopic dependence of the mean field
spin-orbit splitting. The splitting becomes larger for neutrons
than for protons in nuclei having $N>Z$. The general arguments
presented initially (based on the properties of the two-body
spin-orbit and tensor interactions) gave a result in fair agreement
with the empirical observations. A further microscopic study within
the Walecka model supports this initial result, while it was
found that a global fit of Woods-Saxon model parameters appears
to be rather insensitive to the isotopic dependence of the spin-orbit
splitting. A self-consistent calculation using the SIII interaction
gave results in general agreement with the experiment and prediction
by eq.~(3) with negative values of $\beta_{\ell s} \sim -0.6$.

In this context, one should point out that within the Walecka model,
the sign of the isospin term in the spin-orbit potential is in
agreement with the sign of an analogous term  present in the expression
for the central nuclear potential.
While the spin-orbit term in this model is defined, very
approximately, by the $(V-S)$ combination of the entering fields, the
central nuclear potential is proportional to the $(V+S)$ combination.
The main, isoscalar, part of the $(V-S)$ term is positive
and the addition of an isovector contribution, arising from $V^1$, leads
for the $N>Z$ nuclei, as was shown above by us, to an additional
term (positive for neutrons and negative for protons), its  magnitude
growing with $(N-Z)$, together with the ratio of
neutron to proton splittings. At the same time,
the isoscalar part of the central
$(V+S)$ term is negative. The addition  of a $V^1$ term leads
here for neutrons in ($N>Z$) nuclei to reduction of the
absolute value of $(V+S)$. So, with increasing $N$ at a given $Z$,
the depth of the central nuclear potential for neutrons decreases
and they become less bound, while the protons become more
bound. All this is reflected in the W-S model (see eq.~(8) above)
by the fact that $\beta_{\ell s}$ is negative, while $\beta$ is positive.
The two models are thus fully consistent in this respect.

The isotopic dependence of the spin-orbit splitting has also
been studied with methods somewhat different than those used here.
In the work of Mairle \cite{Mairle95} the average spin-orbit potential
was obtained as a convolution with proton and neutron densities
taken in the ratio defined by the short-range two-body spin-orbit
interaction. However the isotopic dependence of the average spin-orbit potential
was not derived here in an explicit form. This point has some
importance, since our analysis, based on the existing empirical data and
different theoretical approaches, resulting in a simple expression,
immediately shows that the difference between the neutron
and proton splittings becomes saturated at large $N$,
which precludes very large
differences. The rather modest difference with a magnitude
of about 10$\%$ seen in the $^{132}$Sn
region is already about $25\%$ of the saturation value, suggesting
that the isospin dependence in itself is unlikely
to lead to dramatic structural
changes. However, in cases of extreme neutron excess,
when the difference between neutron and proton spin-orbit
splittings approaches the maximum value of about 40$\%$
(corresponding to several hundreds of keV) a rather
significant effect on the ordering of levels can be expected.

\section{Acknowledgements}

This work was supported by
the Swedish Natural Research Council,
the Royal Swedish Academy of Sciences
and the Russian Foundation of Fundamental Research (grant No.
00-15-96610).  The authors are grateful to B.L. Birbrair for
discussions.

\newpage

{\bf Table 1. Single particle levels of $^{16}$O.}
\begin{center}

\begin{tabular}{ccccccc}\hline\hline
$n\ell j$&$\varepsilon_{exp}$&Stnd&Set 1&Set 3&SIII-1 &SIII-2\\
\hline\hline
$\nu 1d_{3/2}$&(0.94)&0.89&0.18&0.20&0.66&0.67\\
$\nu 2s_{1/2}$&-3.27&-3.59&-3.89&-3.31&-2.88&-2.87\\
$\nu 1d_{5/2}$&-4.14&-6.97&-6.85&-6.41&-6.87 &-6.89\\
$\nu 1p_{1/2}$&-15.67&-15.06&-16.05&-16.33&-14.58 &-14.56\\
$\nu 1p_{3/2}$&(-21.84)&-19.98&-20.25&-20.10&-20.58 &-20.59\\ \hline
$\pi 1d_{3/2}$&(4.40)&3.76&2.92&3.48&3.55&3.56\\
$\pi 2s_{1/2}$&-0.11&-0.89&-1.14&0.22&0.03&0.03\\
$\pi 1d_{5/2}$&-0.60&-2.76&-2.67&-2.97&-3.57 &-3.59\\
$\pi 1p_{1/2}$&-12.13&-9.95&-10.87&-12.60&-11.17 &-11.15\\
$\pi 1p_{3/2}$&(-18.45)&-14.66&-14.90&-16.40&-17.07 &-17.08\\
\hline\hline
\end{tabular}
\end{center}
{\footnotesize

\noindent
Set 1: $V_0=-52.21$ MeV, $V_{\ell s}=28.6$ MeV $\cdot$ {fm}$^2$,
$a_p=0.67$ fm, $a_n=0.55$ fm  are fixed.

\noindent
Set 3: $V_0=-51.40$ MeV, $V_{\ell s}=25.7$ MeV $\cdot$ {fm}$^2$,
$a_p=0.45$ fm, $a_n=0.50$ fm.}

\vspace{2.0cm}

{\bf Table 2. Single particle states of $^{40}$Ca.}
\begin{center}

\begin{tabular}{ccccccc}\hline\hline
$n\ell j$&$\varepsilon_{exp}$&Stnd&Set 1&Set 3&SIII-1 &SIII-2\\
\hline \hline
$\nu 1f_{5/2}$&-3.48&-2.57&-3.91&-3.54&-1.49 &-1.48\\
$\nu 2p_{1/2}$&-4.42&-3.35&-4.08&-4.69&-2.20 &-2.23\\
$\nu 2p_{3/2}$&-6.42&-5.71&-6.08&-6.57&-4.09 &-4.05\\
$\nu 1f_{7/2}$&-8.36&-10.43&-10.44&-9.72&-9.92 &-9.94\\
$\nu 1d_{3/2}$&-15.64&-16.21&-17.40&-16.43&-15.53 &-15.54\\
$\nu 2s_{1/2}$&-18.11&-16.51&-17.17&-17.00&-15.94 &-15.92\\
$\nu 1d_{5/2}$&-$21.64^*$&-21.08&-21.44&-20.52&-21.90 &-21.90\\ \hline
$\pi 1f_{5/2}$&3.86&4.92&3.79&3.41&4.90&4.91\\
$\pi 2p_{1/2}$&2.64&2.62&2.11&2.07&3.66&3.64\\
$\pi 2p_{3/2}$&0.63&0.89&0.60&0.45&2.23&2.26 \\
$\pi 1f_{7/2}$&-1.09&-2.19&-2.18&-2.85&-3.04 &-3.06\\
$\pi 1d_{3/2}$&-8.33&-7.11&-8.25&-9.01&-8.52 &-8.53\\
$\pi 2s_{1/2}$&-10.85&-8.18&-8.78&-9.30&-8.77 &-8.75\\
$\pi 1d_{5/2}$&-$14.33^*$&-12.05&-12.36&-13.19&-14.74 &-14.75\\
\hline\hline
\end{tabular}
\end{center}
{\footnotesize

\noindent
"Set 1": $V_0=-52.39$ MeV, $V_{\ell s}$ =27.9 MeV $\cdot$ {fm}$^2$;
$a_p=0.67$ fm and $a_n=0.55$ fm are fixed.

\noindent
"Set 3": $V_0=-52.95$ MeV, $V_{\ell s}$ =28.2 MeV $\cdot$ {fm}$^2$,
$a_p=0.63$ fm, $a_n=0.68$ fm.

\noindent
Experimental single particle energy marked by an asterisk
($\ast$) represents a mean value weighted by the
spectroscopic factors.}

\newpage

{\bf Table 3. Single particle states of $^{100}$Sn.}
\begin{center}
\begin{tabular}{ccccccc}\hline\hline
$n\ell j$&$\varepsilon_{sys}$&Stnd&Set 1&Set 3&SIII-1 &SIII-2\\
\hline \hline
$\nu 1h_{11/2}$ &-8.6(5) &-8.66 &-9.01 &-8.72 &-6.35 &-6.87\\
$\nu 2d_{3/2}$ &-9.2(5) &-8.90 &-9.24 &-8.70 &-7.84 &-7.66\\
$\nu 3s_{1/2}$ &-9.3(5) &-9.16 &-9.53 &-9.13 &-7.58 &-7.52\\
$\nu 1g_{7/2}$&-10.93(20)&-11.64&-12.02&-11.23&-10.33 &-9.63\\
$\nu 2d_{5/2}$&-11.13(20)&-11.62&-11.97&-11.59&-10.07 &-10.10\\
$\nu 1g_{9/2}$&-17.93(20)&-17.23&-17.61&-17.21&-16.54 &-17.00\\
$\nu 2p_{1/2}$&-18.38(20)&-19.14&-19.53&-18.93&-19.08 &-18.93\\ \hline
$\pi 1g_{7/2}$&3.90(15)&3.88&3.54&2.70&3.38 &4.04\\
$\pi 2d_{5/2}$&3.00(80)&2.74&2.45&2.64&3.70 &3.69\\
$\pi 1g_{9/2}$&-2.92(20)&-2.01&-2.36&-3.66&-2.74 &-3.16\\
$\pi 2p_{1/2}$&-3.53(20)&-3.48&-3.84&-3.94&-4.80 &-4.65\\
$\pi 2p_{3/2}$&-6.38&-4.95&-5.31&-5.55&-6.22 &-6.18\\
$\pi 1f_{5/2}$&-8.71&-5.54&-5.92&-7.60&-8.43 &-7.89 \\
\hline\hline
\end{tabular}
\end{center}
{\footnotesize

\noindent
Set 1: $V_0=-51.97$ MeV, $V_{\ell s}$ =33.5 MeV $\cdot$ {fm}$^2$;
$a_p=0.67$ fm and $a_n=0.55$ fm  are fixed.

\noindent
Set 3: $V_0=-51.40$ MeV, $V_{\ell s}$ =35.6 MeV $\cdot$ {fm}$^2$,
$a_p=0.52$ fm, $a_n=0.56$ fm.}

\vspace{1.5cm}
{\bf Table 4. Single particle states of $^{132}$Sn.}
\begin{center}

{\footnotesize\begin{tabular}{ccccccccc} \hline \hline
$n\ell j$&$\varepsilon_{exp}$&Stnd&Set 1&Set 2&Set 3&Set 4&
SIII-1& SIII-2 \\ \hline\hline
$\nu 2f_{5/2}$&-0.58&0.36&0.73&0.46&0.22&-0.01&0.67&0.79\\
$\nu 3p_{1/2}$&(-0.92)&-0.13&-0.48&-0.09&-0.55&-0.61&0.16&0.20\\
$\nu 1h_{9/2}$&-1.02&-1.61&-0.84&-1.38&-0.47&-0.97&-0.72&-0.02\\
$\nu 3p_{3/2}$&-1.73&-0.78&-0.88&-0.77&-1.42&-1.32&-0.16&-0.14\\
$\nu 2f_{7/2}$&-2.58&-2.18&-2.55&-2.21&-2.84&-2.52&-1.67&-1.71\\
$\nu 2d_{3/2}$&-7.31&-7.74&-7.45&-7.62&-7.63&-7.77&-8.42&-8.26\\
$\nu 1h_{11/2}$&-7.55&-7.11&-7.96&-7.23&-7.33&-6.60&-7.69&-8.23\\
$\nu 3s_{1/2}$&-7.64&-7.68&-7.73&-7.64&-8.03&-7.93&-8.26&-8.21\\
$\nu 2d_{5/2}$&-8.96&-9.66&-9.94&-9.66&-9.98&-9.69&-10.71&-10.71\\
$\nu 1g_{7/2}$&-9.74&-10.56&-10.04&-10.39&-9.51&-9.81&-11.92&-11.32\\
\hline
$\pi 3s_{1/2}$&(-6.83)&-6.84&-6.87&-6.80&-6.64&-6.70&-4.97&-4.90\\
$\pi 1h_{11/2}$&-6.84&-7.32&-6.66&-7.46&-6.77&-7.48&-5.64&-6.01\\
$\pi 2d_{3/2}$&-7.19&-6.86&-7.20&-6.74&-7.07&-6.72&-5.93&-5.77\\
$\pi 2d_{5/2}$&-8.67&-9.36&-9.20&-9.37&-9.04&-9.30&-7.88& -7.88\\
$\pi 1g_{7/2}$&-9.63&-9.84&-10.41&-9.66&-10.60&-9.81&-10.08 &-9.56\\
$\pi 1g_{9/2}$&-15.71&-14.91&-14.46&-15.00&-14.57&-15.02&-15.03&-15.36\\
$\pi 2p_{1/2}$&-16.07&-16.01&-16.22&-15.92&-16.14&-15.91&-16.68&-16.55\\
\hline \hline
\end{tabular}}
\end{center}
{\footnotesize

\noindent
"Stnd":  $\delta=0.589$ MeV.

\noindent
"Set 1": $V_0=-51.56$ MeV, $V_{\ell s}=33.3$ MeV $\cdot$ {fm}$^2$,
$\beta=1.39$,  $\delta=0.638$ MeV.

\noindent
"Set 2": $V_0=-51.44$ MeV, $V_{\ell s}=34.8$ MeV $\cdot$ {fm}$^2$,
$\beta=1.39$, $\beta_{\ell s}=1.35$, $\delta=0.575$ MeV.

\noindent
"Set 3":  $V_0=-51.55$ MeV, $V_{\ell s}=32.4$ MeV $\cdot$ {fm}$^2$,
$\beta=1.31$, $a_p=0.63$ fm,
$a_n=0.66$ fm, $\delta=0.546$ MeV.

\noindent
"Set 4": $V_0=-51.56$ MeV, $V_{\ell s}=34.1$ MeV $\cdot$ {fm}$^2$,
$\beta=1.34$, $\beta_{\ell s}=1.33$,  $a_p=0.65$ fm,
$a_n=0.66$ fm,  $\delta=0.478$ MeV.

\noindent
Note that some theoretical works \cite{44} postulate that the neutron
$1i_{13/2}$ state at $^{132}$Sn is only 1.9 MeV above the $\nu 2f_{7/2}$
level. Our calculations unequivocally demonstrate, that this state
lies considerably higher, with it's energy equal to +0.55,
+1.59 and +1.02 MeV for the "Stnd", SIII-1
and SIII-2 parameter sets, respectively.}

\newpage
{\bf Table 5. Single particle states of $^{208}$Pb.}
\begin{center}
{\footnotesize\begin{tabular}{ccccccccc} \hline \hline
$n\ell j$&$\varepsilon_{exp}$&Stnd&Set 1&Set 2&Set 3&Set 4&
SIII-1 & SIII-2\\  \hline\hline
$\nu 3d_{3/2}$&-1.40&-0.32&-0.02&-0.23&-0.96&-0.99&0.38&0.42\\
$\nu 2g_{7/2}$&-1.44&-0.79&-0.18&-0.65&-0.89&-1.14&0.01&0.14\\
$\nu 4s_{1/2}$&-1.90&-0.80&-0.70&-0.74&-1.63&-1.51&-0.08&-0.06\\
$\nu 1j_{15/2}$&-$2.09^*$&-2.42&-3.05&-2.31&-2.23&-1.55&-1.41&-1.93\\
$\nu 3d_{5/2}$&-2.37&-1.50&-1.45&-1.40&-2.35&-2.13&-0.39&-0.38\\
$\nu 1i_{11/2}$&-3.16&-4.24&-3.37&-4.05&-2.71&-3.33&-3.37&-2.77\\
$\nu 2g_{9/2}$&-3.94&-3.71&-3.82&-3.59&-4.24&-3.88&-2.91&-2.97\\
$\nu 3p_{1/2}$&-7.37&-7.32&-6.94&-7.17&-7.59&-7.61&-7.21&-7.13\\
$\nu 2f_{5/2}$&-7.94&-8.42&-7.87&-8.25&-8.17&-8.38&-8.59&-8.44\\
$\nu 3p_{3/2}$&-8.27&-8.18&-8.03&-8.04&-8.59&-8.43&-8.18&-8.15\\
$\nu 1i_{13/2}$&-9.00&-9.21&-9.62&-9.08&-8.84&-8.31&-9.73&-10.21\\
$\nu 2f_{7/2}$&-$10.07^*$&-10.57&-10.57&-10.43&-10.72&-10.46&-11.21
& -11.24\\
$\nu 1h_{9/2}$&-10.78&-12.06&-11.35&-11.87&-10.60&-11.09&-13.16&-12.67\\
\hline
$\pi 3p_{1/2}$&$0.17^*$&0.63&0.43&0.72&0.29&0.47&2.79&2.88\\
$\pi 3p_{3/2}$&-0.68&-0.45&-0.46&-0.35&-0.58&-0.69&1.99&2.03\\
$\pi 2f_{5/2}$&-0.97&-0.68&-1.03&-0.60&-1.03&-0.61&0.60&0.74\\
$\pi 1i_{13/2}$&-2.19&-2.86&-2.37&-2.71&-1.94&-2.78&-1.20&-1.53\\
$\pi 2f_{7/2}$&-2.90&-3.38&-3.24&-3.26&-3.21&-3.53&-1.64&-1.66\\
$\pi 1h_{9/2}$&-3.80&-4.60&-5.11&-4.53&-4.71&-4.01&-4.68&-4.24\\
$\pi 3s_{1/2}$&-8.01&-7.76&-7.86&-7.67&-7.87&-7.87&-7.39&-7.33\\
$\pi 2d_{3/2}$&-8.36&-8.41&-8.66&-8.32&-8.59&-8.30&-8.64&-8.51\\
$\pi 1h_{11/2}$&-9.36&-9.33&-8.99&-9.18&-8.60&-9.21&-9.35&-9.65\\
$\pi 2d_{5/2}$&-10.04$^*$&-10.10&-10.05&-9.98&-9.96&-10.15&-10.29
&-10.28\\
$\pi 1g_{7/2}$&-12.18$^*$&-12.07&-12.45&-11.99&-12.08&-11.58&-13.94&-13.59\\
\hline\hline
\end{tabular}}
\end{center}
{\footnotesize

\noindent
The "standard" set of parameters corresponds to $V_0=-51.50$
MeV, $V_{\ell s}=33.2$ MeV $\cdot$ {fm}$^2$, $\beta=\beta_{\ell s}=+1.39$,
$a_p=0.67$ fm, $a_n=0.55$ fm and $\delta=0.604$ MeV.

\noindent
"Set 1" corresponds to $V_0=-51.39$ MeV, $V_{\ell s}=33.1$ MeV $\cdot$
{fm}$^2$, $\beta=1.43$ with $\beta_{\ell s}=-0.6$, $a_p=0.67$ fm,
$a_n=0.55$ fm fixed; $\delta=0.654$ MeV.

\noindent
"Set 2" corresponds to $V_0=-51.34$ MeV, $V_{\ell s}=33.1$ MeV $\cdot$
{fm}$^2$, $\beta=1.40$, $\beta_{\ell s}=1.26$ with $a_p=0.67$ fm,
$a_n=0.55$ fm fixed; $\delta=0.593$ MeV.

\noindent
"Set 3" corresponds to $V_0=-51.99$ MeV, $V_{\ell s}=32.7$ MeV $\cdot$
{fm}$^2$, $\beta=1.36$, $a_p=0.73$ fm,
$a_n=0.72$ fm with $\delta=0.369$ MeV;
$\beta_{\ell s}=-0.6$ is fixed.

\noindent
"Set 4" corresponds to $V_0=-51.93$ MeV, $V_{\ell s}=35.2$ MeV $\cdot$
{fm}$^2$, $\beta=1.38$, $\beta_{\ell s}=1.76$,  $a_p=0.73$ fm,
$a_n=0.72$ fm;  $\delta=0.366$ MeV.

\noindent
Experimental single particle energy marked by an asterisk
($\ast$) represents a mean value weighted by the
spectroscopic factors.}

\vspace{0.3cm}

{\bf Table 6. Magnitudes in MeV of neutron and proton spin-orbit
splittings.}
\begin{center}

\begin{tabular}{cccccccc}\hline \hline
Nucleus&$n\ell j$&$\Delta_{exp}$&Stnd&Set 1&Set 3&SIII-1&SIII-2 \\
\hline \hline

$^{132}$Sn& $\nu 2d$&1.65&1.92&2.49&2.35&2.29&2.45\\
 & $\pi 2d$&1.48&2.50&2.00&1.97&1.95&2.11\\
\\
$^{208}$Pb& $\nu 2f$&2.13&2.15&2.70&2.55&2.62&2.80\\
 & $\pi 2f$&1.93&2.70&2.21&2.18&2.24&2.40\\
\\
 & $\nu 3p$&0.90&0.86&1.09&1.00&0.97&1.02\\
 & $\pi 3p$&0.85&1.08&0.89&0.87&0.80&0.85\\
\hline\hline
\end{tabular}
\end{center}
{\footnotesize

\noindent
Notation is as in previous tables. Data are given only
in the cases where spin-orbit partners of both neutrons
and protons in identical orbits have been observed
experimentally. Note that the splittings are practically
identical for neutrons and protons in the $N=Z$ nuclei,
which are not
included in this Table.}

\end{document}